\documentclass[a4paper,12pt]{article}
\usepackage[utf8]{inputenc}
\pdfoutput=1
\usepackage{jheppub,float}
\usepackage[normalem]{ulem}
\usepackage{xcolor}
\usepackage{amsmath}
\colorlet{shadecolor}{lightgray}
\usepackage{hyperref}
\usepackage{parskip}

\newcommand{\cZ}{\mathcal{Z}}

\newcommand{\tr}{\text{Tr}}

\title{\boldmath A Matrix Model with BMS$_3$ Constraints}
\author{Arindam Bhattacharjee$^{a,b,c}$,}
\author{Neetu$^{d,e}$}
\emailAdd{arindam.bhattacharjee@students.iiserpune.ac.in}
\emailAdd{neetujangid7994@gmail.com}	

\affiliation{$^a$Indian Institute of Science Education and Research Pune \\Dr. Homi Bhabha Road, Pashan, Pune 411 008, India}
\affiliation{$^b$Harish-Chandra Research Institute, Chhatnag Road, Jhunsi, Prayagraj - 211019.}
\affiliation{$^c$ Homi Bhabha National Institute, Training School Complex, Anushaktinagar, Mumbai 400094, India}
\affiliation{$^d$Indian Institute of Science Education and Research Bhopal\\
	Bhopal Bypass Road, Bhauri, Bhopal 462066, India } 
\affiliation{$^e$Dublin Institute for Advanced Studies\\ 10 Burlington Road, Dublin 4, D04 C932, Ireland}

\abstract{Our goal is to find a matrix model with $BMS_3$ constraints built in. These constraints are imposed through Loop equations. We solve them using a free field realisation of the algebra and write down the partition function in eigenvalue form. We comment on the nature of this partition function and it's relations with integrable systems.}

\begin{document} 	
\maketitle
\flushbottom

\section{Introduction}
Although we are far from understanding the complete picture of quantum gravity, matrix models have proven to be very successful in the study of 2D quantum gravity \cite{Klebanov:1991qa, Mukhi:2003sz}. A one-dimensional Hermitian matrix model in the double-scaling limit describes a two-dimensional string theory which can be interpreted as a Liouville theory coupled to $c=1$ matter \cite{GROSS1991459}. This connection generated a huge interest in other possible relations between gravity theories and matrix models. The 2D quantum gravity models are usually formulated as conformal field theories, therefore, one natural direction is to look for a CFT formulation of matrix models \cite{Marshakov:1991gc}. The connection between random matrix models and conformal field theory (CFT) is bilateral. While matrix model techniques can be useful for computing certain correlators in a conformal field theory, in some cases, the techniques of CFT might be useful for solving matrix models.\\
A matrix model possesses an infinite number of symmetries, which gives rise to a recursive relation between correlation functions through Loop equations \cite{Eynard:2015aea}. The existence of infinite symmetries points to a possible integrability structure, which has been extensively explored in the literature \cite{Gerasimov:1990is, Bowick:1990qc, Martinec:1990qg, Alvarez-Gaume:1990asn}. The partition function of matrix models is known to play the role of tau-function of some integrable systems. Their underlying integrability structure makes them exactly solvable and thus, an extremely important tool in the study of lower-dimensional quantum field theories.\\
The loop equations can be reformulated in terms of linear differential constraints on the partition function, where the differential operators satisfy an infinite dimensional algebra \cite{Mironov:1990im, Kostov:1999xi, Fukuma:1990jw,Dijkgraaf:1990rs}. The most famous example is that of Hermitian one matrix model for which the operators are known to satisfy the Virasoro algebra. The matrix model partition function can then be described as a solution to Virasoro-constraints. It is possible to invert this relation and start from the Virasoro algebra (in fact any infinite dimensional algebra) to write a corresponding matrix model partition function. A systematic approach was developed in \cite{Marshakov:1991gc,Kharchev:1992iv} (see \cite{Morozov:1995pb} for review). Their approach gives a formulation of matrix models in terms of conformal field theory, where the constraints imposed on the partition function are translated to conditions imposed on the correlators of a CFT. Our case of interest in this paper, is the BMS$_3$ algebra that arises as the asymptotic symmetry algebra of (2+1)-dimensional flat  space-times \cite{Barnich:2010eb,Barnich:2012aw}.  It is extremely important to understand the behaviour of these algebras from the context of flat space holography \cite{Barnich:2010eb}. While this serves as an example of the method developed in \cite{Marshakov:1991gc}, our bigger motivation is to look for a possible connection between matrix models and higher-dimensional gravity theories. Given the success of matrix models in the study of 2D gravity, we believe that a framework for higher-dimensional theories in terms of matrix models might be useful. A matrix model possessing BMS$_3$ invariance in it's partition function may help us explore the integrability structure that underlines it.\\
We start with a set of linear differential constraints, which we refer to as \emph{BMS$_3$-constraints}. Assuming that those constraints describe the loop equations of a matrix model, we use the free field realisation of BMS$_3$ to write down a solution for those loop equations. Since this methodology is new in the BMS$_3$ literature, we review the basics before using them to compute the desired result. The contents of this paper are organised as follows:
In section \ref{sec:alg-to-mm}, we briefly discuss the method of \cite{Marshakov:1991gc}, to compute a matrix model partition function as a solution to a set of infinite constraints. This will be our guiding principle for this work. In section \ref{sec:BMS3}, we discuss about the BMS$_3$ algebra and its free field realisation. Here we also construct the generators of BMS$_3$ in terms of modes of the free fields. We start section \ref{sec:BMS-MM} by writing the BMS$_3$ generators in terms of a set of oscillator modes. This construction translates the constraints of the algebra into infinite coupled differential equations. Next, we use CFT techniques to write a matrix model partition function in eigenvalue representation which satisfies BMS$_3$ constraints.\\
Our final result is a two-matrix model partition function written in terms of their eigenvalues. It may be interpreted as a $\beta_1 - \beta_2$ matrix model with $\beta_1=-1$ and $\beta_2=-2$ interacting through the measure of the partition function. 

\section{From infinite dimensional algebras to matrix models}
\label{sec:alg-to-mm}
The loop equations for a one-matrix model describing an ensemble, $E$ of $N\times N$ Hermitian matrices, $H\in E$, take the following from
\begin{equation}
\begin{aligned}
\sum_{l=0}^{\mu_1 -1} \langle \tr H^{l}\tr H^{\mu_1 -l -1}\prod_{i=2}^{n}\tr H^{\mu_i}\rangle + \sum_{j=2}^{n} \mu_j\langle\tr H^{\mu_1 + \mu_j -1}\prod_{\substack{i=2\\ i\neq j}}^{n} \tr H^{\mu_i}\rangle\\ = \langle \tr V'(H) H^{\mu_1}\prod_{i=2}^{n}\tr H^{\mu_i} \rangle .
\end{aligned}
\end{equation}
They are an infinite set of recursion relations among the correlation functions that follow from the invariance of the matrix integral under a change of integration variables \cite{Eynard:2015aea}.
The exact loop equations are difficult to solve for finite $N$. However, in the large-$N$ limit, these can be efficiently used to compute correlation functions, order by order in $1/N$ expansion \cite{Migdal:1983qrz}. They also admit a topological expansion and can be formulated as \emph{topological recursion relations} among the correlation functions \cite{Eynard:2004mh,Eynard:2007kz}.\\
For a one-matrix model, the conjugation of ensemble elements by unitary matrices acts as a gauge symmetry on the partition function, with the most general form of the potential given by
\begin{equation}\label{eq:potential}
V(H) = -\sum_{k=0}^{\infty} t_{k}H^{k} .
\end{equation}
The correlators are obtained as derivatives of the partition function, with respect to the parameters $t_k$,
\begin{equation}\label{eq:t_derv}
\begin{aligned}
\langle \tr H^{\mu_1}\cdots \tr H^{\mu_n}\rangle
= \frac{\partial^n}{\partial t_{\mu_1}\cdots \partial t_{\mu_n}}\cZ .
\end{aligned}
\end{equation}
This relation can be used to rewrite loop equations as linear differential constraints,
\begin{equation}\label{eq:Virasoro_constraints}
L_{n}\cZ = 0 \quad \text{for} \quad n\geq -1,
\end{equation}
where
\begin{equation}\label{eq:Ln-1MM}
L_{n} = \sum_{k=0}^{\infty} k t_{k}\frac{\partial}{\partial t_{n+k}} + \sum_{k=0}^{n}\frac{\partial}{\partial t_{k}}\frac{\partial}{\partial t_{n-k}}.
\end{equation}
The operators $L_n$ satisfy a closed algebra, 
\begin{equation}\label{eq:VA}
[L_n, L_m] = (n-m)L_{n+m}, 
\end{equation}
which is similar to the Virasoro algebra, except that $n,m\geq -1$. Therefore, it is referred to as ``discrete Virasoro algebra". The constraints (\ref{eq:Virasoro_constraints}) along with
\begin{equation}
\frac{\partial}{\partial t_0} \cZ = N \cZ ,
\end{equation}
are called the \emph{Virasoro constraints}. The matrix model partition function is given as a solution of these infinite set of constraint equations.\\
A formulation of the matrix model partition function as a solution to Virasoro constraints points to a connection between Hermitian one matrix model and 2d CFT \cite{Kostov:1999xi}. In \cite{Marshakov:1991gc}, a systematic approach was developed to construct solutions of such constraints \cite{Mironov:1990im,ambjorn1990multiloop}, using the methods of conformal field theory.\\
The idea is to identify the operators, $L_n$ with the modes of stress tensor, $T_n$ of a conformal field theory. The solution to differential constraints is then obtained as a correlator in the CFT, and the annihilation of that correlator by $L_n$ is translated to the annihilation of vacuum state by $T_n$. Thus, finding an integral expression of the partition function essentially involves two main steps:
\begin{itemize}
\item [(i)] Finding a $t$-dependent ``Hamiltonian" operator that relates $L_n$ with the modes of stress tensor of a CFT. The identification is expressed through
\begin{equation}\label{eq:step-1}
L_n \langle N|e^{H(t)} = \langle N|e^{H(t)} T_n . 
\end{equation}
where $\langle N| $ is a charged vacuum state of the theory.

\item [(ii)] Finding states $|G\rangle $ in the CFT, which are annihilated by $T_n$, $n\geq -1$,
\begin{equation}
T_n |G\rangle = 0.
\end{equation}
\end{itemize}
Once we find $H(t)$ and $|G\rangle $, the solution is given by
\begin{equation}\label{eq:solution}
\cZ = \langle N| e^{H(t)}|G\rangle .
\end{equation}

The construction of operator $H(t)$ is somewhat ad hoc. This Hamiltonian operator, in general, does not have any relation to the CFT Hamiltonian, since there is no obligation for a CFT Hamiltonian to satisfy such a relation. However, the state $|G\rangle $ is well-known. It is given by the action of an operator $G$ which commutes with the stress tensor, on the uncharged vacuum state,
\begin{equation}\label{eq:TG_comm}
|G\rangle = G |0\rangle, \quad \text{where}\quad [T_n, G] = 0, \quad n\geq -1 
\end{equation}
Any function of the screening charges satisfy such a commutation relation. In a CFT, screening charges are operators with non-zero ``charge" under the conserved current but zero conformal dimension. Thus, adding them to a correlation function will not change their conformal behaviour, however, it will change their total charge.\\
A solution (\ref{eq:solution}) can be constructed for any algebra of constraints, provided, the following three conditions are fulfilled:
\begin{itemize}
	\item The algebra admits a free field realisation.
	\item One can find a vacuum annihilated by relevant generators in the corresponding field theory.
	\item One can find a free field representation of the screening charges.
\end{itemize}
It is well known that a free field realisation of the Virasoro algebra is given by a free bosonic CFT, and a solution to the Virasoro constraints can be constructed systematically using the above mentioned procedure. There exist generalisations to other set of constraints, called the \emph{W-constraints}, whose solutions correspond to multi-matrix models. The solution of $W_{r+1}$-constraints,
\begin{equation}
W_{n}^{(a)}Z = 0, \quad n\geq 1-a, \quad a=2,\cdots r+1,
\end{equation}
where $W_{n}^{(a)}$ satisfy $W_{r+1}$-algebra, is given by an $r$-matrix integral \cite{Mironov:1990im,ambjorn1990multiloop}, with $r$ being the rank of the algebra. In this case, the associated CFT is that of $r$ free scalar fields. For a construction of the solution to Virasoro and $W$-constraints, refer to \cite{Marshakov:1991gc,Mironov:1990im,ambjorn1990multiloop}.
In this paper, we adopt this construction for the case of BMS$_3$ algebra, which is the asymptotic symmetry algebra of 3D flat spacetimes.

\section{The BMS$_3$ algebra}\label{sec:BMS3}
In (2+1) dimensions, pure gravity has no propagating degrees of freedom which makes all the solutions of Einstein's equation locally equivalent. The only non-trivial features are global and it can be shown that with proper boundary conditions on metric components, an infinite number of degrees of freedom can live at the boundary. These boundary conditions specify the asymptotic phase space of the theory, and BMS$_3$ is the symmetry associated with it. This makes the problem of finding a matrix model partition function with BMS$_3$ invariance extremely interesting.\\
In 3D, the oldest set of boundary conditions on the metric was given by Brown and Henneaux \cite{Brown:1986nw} for asymptotically AdS$_3$ spacetimes. The flat limit of these conditions were studied by Barnich et al \cite{Barnich:2012aw, Barnich:2013jla}. They take the 3D analogue of BMS ansatz,
\begin{equation}
	ds^2 = e^{2\beta}\frac{V}{r} du^2 - 2 e^{2\beta}dudr+r^2 (d\phi - Udu)^2 ,
\end{equation}
which was originally put forward in four dimensions\cite{Bondi:1962px, Sachs:1962wk}. The fall-off conditions are then specified as the large $u$ behaviour of the functions $V,\beta,U$ all of which are, at this point, arbitrary function of the co-ordinates $\{u,r,\phi\}$. The chosen fall-off are
\begin{align}\label{BMS3_falloff}
	\frac{V}{r} = O(1); \qquad \beta = O(1/r); \qquad U = O(1/r^2).
\end{align}
Demanding that these also satisfy Einstein's equation gives the form of the asymptotic metric for our case,
\begin{equation}\label{asymp_flat_metric}
	ds^2 = \mathcal{M} du^2 - 2 du dr + 2\mathcal{N}dud\phi + r^2 d\phi^2 ,
\end{equation}
where $\mathcal{M},\mathcal{N}$ are two arbitrary functions of $\{u,\phi\}$ that span the asymptotic phase space.
The asymptotic symmetries would be the infinitesimal transformations that keep the above form of the metric unchanged. Thus, we solve for the asymptotic killing vector field, $\xi$ that satisfies
\begin{align}
	&\mathcal{L}_{\xi} g_{rr} = 0 \qquad \mathcal{L}_{\xi} g_{r\phi} = 0 \qquad \mathcal{L}_{\xi} g_{\phi\phi} = 0\nonumber\\
	&\mathcal{L}_{\xi} g_{ur} = o(1/r) \qquad \mathcal{L}_{\xi} g_{uu} = o(1) \qquad \mathcal{L}_{\xi} g_{u\phi} = o(1).
\end{align}
This vector field equipped with a modified Lie bracket gives the BMS$_3$ algebra.
BMS$_3$ algebra is spanned by two spin-2 fields, $T_n$ and $M_n$ and their commutators are given by
\begin{align}\label{bms3_alg}
	[T_n,T_m] &= (n-m) T_{n+m} + \frac{c_1}{12} n (n^2 -1) \delta_{n+m,0}, \nonumber\\
	[T_n,M_m] &= (n-m) M_{n+m} + \frac{c_2}{12} n (n^2 -1) \delta_{n+m,0},\\
	[M_n,M_m] &= 0. \nonumber 
\end{align} 
\subsection{Free field realisation of BMS$_3$}

As it turns out, there exists a free field realisation of this algebra in terms of the $\beta-\gamma$ bosonic ghost CFT. It was shown \cite{Banerjee:2015kcx} that a twisted ghost system with spin (2,-1) of the fields, respectively, can realise the above algebra (\ref{bms3_alg}). 

The bosonic $\beta-\gamma$ system (see \cite{Friedan:1985ge}) generically has a field $\beta$ with spin $\lambda$ and $\gamma$ with spin $1-\lambda$ and they satisfy the following OPE,
\begin{equation}\label{bg_ope}
\gamma(z) \beta(w) \sim  \beta(w) \gamma(z) \sim \frac{1}{(z-w)},
\end{equation}
while the OPE among the fields with themselves vanishes.
Our interest lies in the case where $\lambda = 2$. For this system, the holomorphic parts of the primary fields can be expanded as
\begin{equation}\label{eq:bg_mode}
\beta(z)=\sum_{n\in \mathbb{Z}}\beta_n z^{-n-2}, \quad \gamma(z)=\sum_{n\in\mathbb{Z}}\gamma_n z^{-n+1},
\end{equation}

The stress tensor of the theory,
\begin{align}\label{eq:bg_T}
T(z)&=-\lambda:\beta(z)\partial_z\gamma: + (1-\lambda):\gamma(z)\partial_{z}\beta:\\
&= -2:\beta(z)\partial_z\gamma: - :\gamma(z)\partial_{z}\beta:\quad \text{(for $\lambda$ = 2)},
\end{align}	
has the mode expansion
\begin{equation}
T(z)=\sum_{n\in\mathbb{Z}}T_{n} z^{-n-2}, 
\end{equation}
with
\begin{equation}
T_n = \sum_{m=0}^{\infty}(2n+m)\beta_{-m}\gamma_{m+n} + \sum_{m=0}^{\infty} (n-m)\gamma_{-m}\beta_{m+n} + \frac{1}{2}\sum_{\substack{a+b=n\\a,b\geq 0}}(2a + b)\beta_{a}\gamma_b
\end{equation}
From the OPE (\ref{bg_ope}), it is clear that the above stress tensor satisfies the following expansions:
\begin{align}\label{T_beta_OPE}
T(z) T(w) &\sim \frac{1}{2} \frac{26}{(z-w)^4} + \frac{2T(w)}{(z-w)^2} + \frac{\partial T(w)}{(z-w)}\\
T(z) \beta(w) &\sim \frac{2\beta(w)}{(z-w)^2} + \frac{\partial \beta(w)}{(z-w)}
\end{align}
This suggests that the modes of $T(z)$ and $\beta(z)$ satisfy an algebra that is almost like BMS$_3$ except that the central charges are different. The central charges of BMS$_3$ (\ref{bms3_alg}) are arbitrary, whereas in this system $c_1 = 26$ and $c_2 = 0$. To get around this problem, \cite{Banerjee:2015kcx} twisted the above stress tensor with
\begin{equation}
T(z) \rightarrow T(z) - a \partial^3 \gamma
\end{equation}
This twist introduces an arbitrary central charge of $12a$ in the OPE of $T(z)$ with $\beta(z)$.  So now we may say that the modes of the stress tensor gives the $T_n$ generator whereas the modes of the field $\beta$ acts as $M_n$. Of course, the $c_1$ central charge is still fixed to be 26. But we can change that by coupling this system with arbitrary chiral matter whose stress tensor has some non-zero central charge. \textit{We ignore this extra complication for now as whatever we derive with the twisted $\beta-\gamma$ system described above would go through even in that case.}

\section{A BMS$_3$ invariant Matrix Model}\label{sec:BMS-MM}
We impose an infinite set of differential constraints
\begin{equation}\label{eq:BMS-const}
B_{n}^{a} Z = 0, \quad n\geq -1, a=1,2.
\end{equation}
such that the operators $B_n^{a}$ satisfy BMS$_3$ algebra (\ref{bms3_alg}). The explicit form of these operators is given in (\ref{eq:BMS_Ln}) and (\ref{eq:BMS_Mn}).  We call these constraints the BMS$_3$-constraints, and claim that a solution of (\ref{eq:BMS-const}) gives a BMS$_3$ invariant matrix model partition function. The constraints (\ref{eq:BMS-const}) should describe the loop equations of that matrix model.

\subsection{Loop equations}
In order to define the differential operators corresponding to Loop equations, we first observe that the OPE (\ref{bg_ope}) gives the following relation between the modes:
\begin{align}\label{com_mode}
[\gamma_n , \beta_m] = \delta_{n+m,0},
\end{align}
while the rest of the commutators vanish. Thus, the pair $\{\gamma_k,\beta_{-k}\}$ behaves like creation and annihilation operators of a simple harmonic oscillator (SHO) and can be equated with $\{\frac{\partial}{\partial t_k},t_k\}$ for $k>0$. A similar set of equivalence can be made for $k<0$ modes, with another set of SHOs $\{\frac{\partial}{\partial \bar{t}_k},\bar{t}_k\}$. \footnote{A semi classical oscillator construction of BMS$_3$ algebra was also done in \cite{Ammon:2020wem}. We thank Shouvik Datta for pointing out this reference.}

Thus, the differential operators of relevance become
\begin{eqnarray}
\label{eq:BMS_Ln}\nonumber
B_n^{1}\equiv L_n &=& \sum_{m=0}^{\infty}(2n+m)t_{m}\frac{\partial}{\partial t_{m+n}} + \sum_{m=0}^{\infty} (m-n)\bar{t}_{m}\frac{\partial}{\partial \bar{t}_{m+n}}\\ &+& \frac{1}{2}\sum_{\substack{a+b=n\\a,b\geq 0}}(2a + b)\frac{\partial}{\partial t_a}\frac{\partial}{\partial\bar{t}_{b}},
\\
\label{eq:BMS_Mn}
B_n^{2}\equiv M_n &=& \frac{\partial}{\partial\bar{t}_n}, \quad n>0, \quad M_n = t_{-n}, \quad n<0,
\end{eqnarray}
which satisfy
\begin{eqnarray}
\nonumber [ L_n,L_m ]&=&(n-m)L_{n+m},\\
\nonumber \left[ L_n,M_m\right] &=& (n-m) M_{n+m},\\
\left[M_n,M_m\right]&=&0.
\end{eqnarray}
This is the classical version of BMS$_3$ algebra.
The matrix model partition function is obtained as a solution to the constraints
\begin{equation}\label{eq:constraints}
L_n Z_{N} = 0, \quad\text{and} \quad M_n Z_{N} =0 \quad\text{for}\quad n\geq -1,
\end{equation}
where the suffix $N$ of the partition function indicates the charge of the vacuum under the $U(1)$ current of the system, 
\begin{equation}\label{u1_current}
j(z) = - :\gamma \beta:
\end{equation}

The rest of this paper is devoted to finding a solution to (\ref{eq:constraints}).
To solve the above constraints, we will use our knowledge of the free field realisation of the algebra. Our objective would be to translate the constraints in terms of a CFT correlator in the $\beta-\gamma$ model.

\subsection{The `Hamiltonian' function}
The procedure to find the partition function that satisfies (\ref{eq:constraints})  involves two vital steps \cite{Morozov:1995pb}. Firstly, we find a ``Hamiltonian" operator that relates the differential operators with the modes of the operators of our CFT. 

This relation is expressed in terms of the following expressions:
\begin{equation}\label{eq:constraints_explicit}
L_n\langle N|e^{H(t,\bar{t})} = \langle N |e^{H(t,\bar{t})} T_n, \quad \text{and} \quad M_n\langle N|e^{H(t,\bar{t})} = \langle N|e^{H(t,\bar{t})} \beta_n.
\end{equation} 
The state $\langle N |$ is a vacuum of the theory which is charged.\\

We propose the following operator:
\begin{equation}\label{eq:hamiltonian}
H(t,\bar{t}) = \sum_{k > 1}t_k \gamma_{k} + \sum_{k\geq -1}\bar{t}_{k} \beta_{k} 
\end{equation}
Using (\ref{eq:BMS_Ln}), we have
\begin{equation}
L_n \langle N| e^{H(t,\bar{t})} = \langle N| \left(\sum_{p = 0}^{\infty} (2n+p) t_p \gamma_{n+p} + \sum_{p=0}^{\infty} (p-n)\bar{t}_p \beta_{n+p} + \frac{1}{2}\sum_{a+b = n;a,b\geq 0} \gamma_a \beta_b \right),
\end{equation}
whereas to work out the RHS of (\ref{eq:constraints_explicit}), we make use of the BCH formula
\begin{equation}
e^X Y = Y e^X + [X,Y],
\end{equation}
and also use the fact that the vacuum state $\langle N|$ is annihilated by the modes $\beta_{-k},\gamma_{-k}$ for $k>0$. It can be easily shown that the first equation of (\ref{eq:constraints_explicit}) is satisfied.\\
We also need to check the second equality in (\ref{eq:constraints_explicit}). This is relatively easy to check since the LHS gives
\begin{equation}
M_n \langle N| e^{H(t,\bar{t})} = \langle N| \beta_n e^{H(t,\bar{t})}.
\end{equation} 
This result is also obtained from RHS quite straightforwardly as $\beta_n$ commutes with the Hamiltonian for $n>0$ and hence, our choice of Hamiltonian function is justified.
\subsection{Screening Charges}
To complete our analysis, we now require a ket state $|G\rangle$ such that
\begin{equation}\label{ket_constraint}
T_n |G\rangle = 0 \qquad M_n |G\rangle = 0
\end{equation}
If such a state is found, then we may claim that the full partition function of the theory is given by
\begin{equation}\label{partition_fn_form}
Z_N = \langle N| e^{H(t,\bar{t})}|G\rangle,
\end{equation}
which from the properties of (\ref{eq:constraints_explicit}) and (\ref{ket_constraint}) satisfies all our constraints.\\
Generic states, which commute with all positive modes of stress tensor (first equation of (\ref{ket_constraint})) are generated by Screening Operators. 
Unfortunately, for a $\beta-\gamma$ system of spin (2,-1) we don't have a spin 1 primary at hand. Hence, the construction of these operators isn't straightforward. For this we need to take an indirect route which we chalk out below.

For a generic bosonic $\beta-\gamma$ system of dimension $(\lambda,1-\lambda)$ the system has a background charge $(1-2\lambda)$, which implies that our system has a background charge $-3$. The presence of a background charge makes the $U(1)$ symmetry anomalous and the current (\ref{u1_current}) is no longer a primary of the theory.\\
To get a better handle in the theory, we fermionize the theory \cite{Friedan:1985ge}. We take a free scalar field, $\phi$ which satisfies the OPE
\begin{equation}
\phi(z) \phi(w) \sim -\ln (z-w),
\end{equation}
and two fermionic fields, $\eta$ and $\xi$ such that $\eta(z)$ and $\partial\xi(z)$ are primary fields of dimension one. Their OPE is given by
\begin{equation}
\eta(z) \xi(w) \sim \frac{1}{(z-w)}.
\end{equation}
Then in terms of these fields, we can write
\begin{equation}\label{eq:bosonization}
\beta(z) = e^{-\phi(z)}\partial\xi , \quad \gamma(z)=e^{\phi(z)}\eta(z).
\end{equation}
This map gives us incredible advantage. Since, we know that free scalar fields have Vertex operators 
\begin{align}
\mathcal{V}_{\alpha} (z,\bar{z}) = e^{i\sqrt{2}\alpha \phi(z,\bar{z}) } ,
\end{align}
which are primary operators with dimension\footnote{In absence of a background charge. Otherwise $h_{\mathcal{V}_{\alpha}} = \alpha^2 - 2\alpha_0 \alpha$} $\alpha^2$. Hence, we can construct dimension one primaries now, which in turn gives us our screening charges. Of course we also have fermionic primaries in our theory. It turns out the relevant screening charges in this new theory are 
\begin{equation}
Q_1 = \oint e^{-\phi(z)}, \quad Q_2=\oint e^{-2\phi(z)},
\end{equation}
as $T_n$ and $M_n$ both commute with them\footnote{This is the reason we couldn't use the fermionic screening operators. $M_n$ doesn't commute with them.}. It also implies that we can take any function of these charges acting on vacuum as our state $|G\rangle$. 
\subsection{The matrix model partition function}
If we choose $G$ to be an exponential function, then we realise that charge conservation of CFT correlators demand that only the N-th term of the exponential operator would survive. Thus, the final partition function is given by
\begin{align*}\label{partition_fn_subs}
	Z_N &= \langle N| e^{H(t,\bar{t})}|G\rangle,\\
	&= \frac{1}{N_1 ! N_2 !}\langle N|e^{H(t,\bar{t})}(Q_1)^{N_1}(Q_{2})^{N_2}|0\rangle, \qquad \text{with $N_1 + N_2 = N$}.
\end{align*}
To evaluate the above integral, we first write our Hamiltonian in terms of fields,
\begin{equation}\label{eq:hamiltonian}
H(t,\bar{t}) = \sum_{k > 1}t_k \gamma_{k} + \sum_{k\geq -1}\bar{t} \beta_{k} = -\oint V(z)\partial\gamma(z)-\oint U(z)\partial\beta(z),
\end{equation}
where
\begin{equation}\label{eq:UV}
V(z)=\sum_{k>1} t_k \frac{z^{k-1}}{k-1}, \quad U(z)=\sum_{k\geq -1}\bar{t}_k \frac{z^{k+2}}{k+2}.
\end{equation}
Thus, in terms of fields our partition function becomes
\begin{align*}
Z_{N}&=&\frac{1}{N_1 ! N_2 !}\langle N| :e^{-\oint V(z)\partial\gamma(z)-\oint U(z)\partial\beta(z)}:\prod_{i=1}^{N_1}\oint_{C_i}dx_{i}:e^{-\phi(x_{i})}:\prod_{j=1}^{N_2}\oint_{C_j}dy_{j}:e^{-2\phi(y_{j})}:|0\rangle
\end{align*}
We'll also need the OPE relations between the original fields and the new fields, which are
\begin{eqnarray}
\partial\beta(z)\phi(z')&\sim& \frac{\beta(z')}{z-z'} ,\\
\partial\gamma(z)\phi(z')&\sim& -\frac{\gamma(z')}{z-z'} .
\end{eqnarray}
Finally, to evaluate the correlator, we use the identity of exponentiated operators
\begin{equation}\label{eq:exp_identity}
\langle :e^{A_1}::e^{A_2}:....:e^{A_n}:\rangle = \text{exp}\sum_{i<j}^{n}\langle A_i A_j\rangle .
\end{equation}
Using these, we get
\begin{equation}\label{eq:final_Z}
Z_{N}=\frac{1}{N_1 ! (N-N_1) !}\prod_{i=1}^{N_1}\oint_{C_i}dx_{i}\ e^{X(x_i)}\prod_{j=1}^{(N-N_1)}\oint_{C_j}dy_{j}\ e^{Y(y_j)}\frac{1}{\triangle(x)\triangle^{4}(y)\triangle^{2}(x,y)},
\end{equation}
where
\begin{eqnarray}
X(x_i) = V(x_i)\gamma(x_i)-U(x_i)\beta(x_i),  \quad Y(y_j)=V(y_j)\gamma(y_j)-U(y_j)\beta(y_j)\\
\vspace{5pt}
\triangle(x)=\prod_{i<k}^{N_1}(x_{i} - x_k), \quad \triangle(y)=\prod_{j<k}^{N_2}(y_{j} - y_k), \quad \triangle(x,y)=\prod_{i,j}(x_i - y_j)
\end{eqnarray}
This is our final result.\\
A $\beta$-matrix model is defined for any complex $\beta$ with the integral measure
\begin{equation}
dM = |\triangle(\Lambda)|^{\beta}d\Lambda\ dU_{Haar},
\end{equation}
for an ensemble $E_{N}^{\beta}$ of $N\times N$ matrices $M$. $d\Lambda$ is the measure on eigenvalues and $dU_{Haar}$ defines the measure on the corresponding circular ensemble (diagonalising matrices). The values $\beta = 1,2,4$ correspond to an ensemble of real symmetric matrices, complex Hermitian matrices and quaternionic Hermitian matrices, respectively. Our matrix model (\ref{eq:final_Z}) appears to be a hybrid two matrix model with $\beta_1 = -1$ and $\beta_2 = -4$, interacting through the measure.

\section{Discussion}
In this work, we have found a matrix model partition function with $BMS_3$ constraints. The $BMS_3$ constraints are imposed through loop equations which suggests it might also be possible to formulate a topological recursion relation. In recent times BMS symmetry has appeared in many different contexts. For example, in the description of tensionless strings\cite{Bagchi:2013bga} and in non-relativistic holography\cite{Bagchi:2009my}. Following these, there has been a lot of work trying to understand the possible connections between $BMS_3$ algebras and integrability \cite{Fuentealba:2017omf}. It is natural to expect the partition function given in (\ref{eq:final_Z}) sheds some light in this direction. In case of the hermitian matrix models, the recursion relations turn out to be the $\tau$ functions of some well-known integrable systems (see \cite{Marshakov:1993au} and references therein). Whether our constraints also reduce to a similar form will be an interesting study. In particular, the relation between these constraints and the integrable system described in \cite{Fuentealba:2017omf} would require further investigations.\\
$BMS_3$ algebra is also linked with flat limits of Liouville theory \cite{Barnich:2012rz} and it would be interesting to understand the connection between this 2D gauge theory and our resulting matrix model.\\
Another possible direction would be to understand Super-BMS$_3$ algebras \cite{Barnich:2014cwa, Fuentealba:2017fck, Banerjee:2018hbl, Banerjee:2019lrv} in terms of matrix models. They appear as the asymptotic symmetry group of Supergravity theories in 3D flat spacetimes and their free field realisation were discussed in \cite{Banerjee:2016nio}. The non-trivial features resulting from fermionic constraints on a Matrix model partition function would be interesting to study.
\section*{Acknowledgements}
We would like to thank Nabamita Banerjee, Arghya Chattopadhyay, Suvankar Dutta and Dileep Jatkar for their valuable comments on our draft. We would also like to thank Subhroneel Chakroborty and Koushik Ray for suggesting useful references. AB would like to thank IISER Bhopal for their hospitality during part of this project. Finally, we thank the people of India for their continuous support for basic sciences.

\bibliographystyle{ieeetr}
{\small
	\bibliography{references}}

\end{document}